# First Measurements of Electron Temperature Fluctuations by Correlation ECE on Tore Supra

V.S. Udintsev, M. Goniche, J.-L. Ségui, G.Y. Antar[1],
D. Molina, G. Giruzzi, A. Krämer-Flecken[2], and the Tore Supra Team

*Association Euratom-CEA, CEA/DSM/DRFC, CEA/Cadarache,
F-13108 St. Paul-lez-Durance, France*
[1] *Center for Energy Research, UCSD, 9500 Gilman Dr., La Jolla CA 92093, USA*
[2] *Association Euratom-FZJ, IPP Forschungzentrum Jülich GmbH, Germany*

ABSTRACT. Electron temperature fluctuation studies can help to understand the nature of the turbulent transport in tokamak plasmas. At Tore Supra, a 32-channel heterodyne ECE radiometer has been upgraded with two channels of 100 MHz bandwidth and tunable central frequencies allowing the shift of the plasma sample volume in the radial direction. With the sufficiently large video bandwidth and the long sampling time, it is possible to reduce significantly the thermal noise and to identify "true" high frequency components up to 200 kHz from the cross-correlation between these channels. First results of temperature fluctuation measurements on Tore Supra are reported in this paper.

## 1. Introduction

Studies of the plasma turbulence aid in understanding the nature of the transport properties in fusion. Two general types of fluctuations, electrostatic and magnetic, can be distinguished in tokamak plasmas. Magnetic fluctuations brake the nested flux surfaces and, therefore, enhance the plasma transport. Electrostatic fluctuations do not destroy the nested magnetic topology, however, the enhancement of transport is due to the $\vec{E} \times \vec{B}$ drifts from the fluctuating electric fields [1]. In order to establish a relation between the fluctuations and transport, measurements of various fluctuating components, such as $\tilde{n}_e$, $\tilde{T}_e$, $\tilde{B}_r$ and $\tilde{E}_\theta$, as well as knowledge of correlations between them, are needed.

Measurements of electron temperature fluctuations ($\tilde{T}_e$) by means of correlation Electron Cyclotron Emission (ECE) diagnostics are a relatively new subject in probing the nature of the turbulent transport in fusion plasmas. Results obtained at TEXT-U [2 - 4], W7-AS [5], RTP [6], TEXTOR [7, 8] and Alcator C-Mod [9] have yielded some interesting information on microturbulence properties both of electrostatic and magnetic origins. However, no clear characteristics of the turbulent fluctuations in different operational regimes of tokamak plasmas have been obtained up to present days. These measurements generally require long integration times. Therefore, they will find their best application in experiments in which long(several seconds) stationary plasmas are attained.

On Tore Supra tokamak ($R_0$ = 2.40 m, $a$ = 0.72 m, $B_T$ < 4 T, circular cross-section), a 32-channel heterodyne ECE radiometer has recently been upgraded to include two channels for temperature fluctuation measurements with a radial resolution of about 1 cm. Experiments have been performed for about three weeks during the 2004 campaign. The experimental setup, as well as the first results, are presented in this paper.

## 2. Theoretical background

For optically thick plasmas (optical thickness $\tau_n \gg 1$, $n$ is a harmonic number), ECE signal $S_{ECE}(t)$ consists of an average $\overline{S}_{ECE}$ and a fluctuation part $\tilde{S}_{ECE}(t)$. In their own turn, these quantities are proportional to the average plasma temperature $\overline{T}_e$ and to its fluctuating component $\tilde{T}_e$ plus the thermal (or photon) noise $\tilde{N}$ (see also Eq. 3 later on in this paper). In general, the measured signal results from the integration over three spatial coordinates for the sample plasma volume: radial $r$, poloidal $\theta$ and toroidal $\varphi$:

$$\tilde{S}_{ECE}(t) = \int_{\Delta f} dr \int_\theta d\theta \int_\varphi d\varphi \tilde{S}'(r, \theta, \varphi, t) + \tilde{S}_{instrum}. \tag{1}$$

Here, $\tilde{S}_{instrum}$ is the 'instrumental' noise caused by video detectors and video amplifiers. The integration on 'real' fluctuation component $\tilde{T}_e$ gives:

$$\tilde{T}_e = \int_{\Delta f} dr \int_\theta d\theta \tilde{T}_e'(r, \theta, t), \tag{2}$$

implying no toroidal ($\varphi$) dependence of $\tilde{T}_e$. One should be aware that the poloidal and radial integration filters the measured data. Therefore, using the raw signals is not impossible but the interpretation is rather difficult. This might have been a limitation in the past as most of the analyses were done using the power spectrum [2].

However, the 'instrumental' noise $\tilde{S}_{instrum}$ produced at the Intermediate Frequency (IF) stage and/or by video detectors may be a (non-linear) function of the input signal, which makes the identification of the 'real' temperature fluctuations much more complicated. Therefore, verification that the 'instrumental' noise does not dominate the true fluctuation spectra is very important in correlation technique.

In many correlation ECE experiments [2 - 7, 9], the following simplified analysis of data to retrieve the information on temperature fluctuations has been used. In this analysis, the 'instrumental' noise component $\tilde{S}_{instrum}$ is assumed to be insignificant and/or totally uncorrelated. ECE signal $S_{ECE}(t)$ from the plasma sample volume can be written in a simple form as:

$$S_{ECE}(t) = \overline{S}_{ECE} + \tilde{S}_{ECE}(t) = c(\overline{T}_e + \tilde{T}_e + \tilde{N}) = c\overline{T}_e \left(1 + \frac{\tilde{T}_e(t)}{\overline{T}_e} + \frac{\tilde{N}(t)}{\overline{T}_e}\right), \tag{3}$$

where $c$ is a proportionality (or calibration) factor for a given ECE signal. From Eq. (3), an expression for the normalized fluctuation component can be written as:

$$\tilde{S}_{ECE}(t) = \frac{S_{ECE}(t) - c\overline{T}_e}{c\overline{T}_e} = \frac{\tilde{T}_e(t)}{\overline{T}_e} + \frac{\tilde{N}(t)}{\overline{T}_e}. \tag{3a}$$

The thermal noise $\tilde{N}$ has nothing to do with the "real" temperature fluctuations $\tilde{T}_e$, therefore, it is necessary to reduce its influence in order to determine "true" fluctuations can be determined. This can be achieved by cross-correlation between two ECE signals whose temperature fluctuations are correlated while the noise is uncorrelated, or even by autocorrelation for a single ECE signal (if the video bandwidth $B_V$ of the ECE radiometer is much larger than the spectral width of temperature fluctuations) [5, 8]. The scheme in which separate frequencies are coming from the same sample plasma volume to observe coherent temperature fluctuations and to decorrelate the thermal noise, is shown in Fig. 1(*a*). If two spatial volumes are different (non-overlapping), the cross-correlation analysis is possible if fluctuations propagate in the plasma (Fig. 1(*b*)). The phase velocity $\vec{v}_{ph}$ and the distance between volumes *r* determine the time delay $\tau_r$:

$$\tau_r = \frac{r}{\vec{v}_{ph}}. \tag{4}$$

From the cross-correlation and the cross-phase between two spatially separated channels, correlation lengths, the wave-numbers and, finally, the dispersion relations $\vec{k}(f)$ can be obtained [10]:

$$k(f)r = 2\pi f \tau_r. \tag{5}$$

The cross-correlation function over time period *T* for two normalized fluctuating components, $\tilde{S}_1(t)$ and $\tilde{S}_2(t)$, and for a given time lag $\tau$, can be written as [11]:

$$R_{12}(\tau) = \frac{1}{T}\int_0^T \tilde{S}_1(t)\tilde{S}_2(t+\tau)dt. \tag{6}$$

For $\tau = 0$ (zero time lag), Eq. (6) takes the following form:

$$R_{12}(0) = \overline{\tilde{S}_1(t)\tilde{S}_2(t)} = \frac{\overline{\tilde{T}_{e1}(t)\tilde{T}_{e2}(t)}}{\overline{T}_{e1}\overline{T}_{e2}} + \frac{\overline{\tilde{N}_1(t)\tilde{N}_2(t)}}{\overline{T}_{e1}\overline{T}_{e2}} + \frac{\overline{\tilde{T}_{e1}(t)\tilde{N}_2(t)}}{\overline{T}_{e1}\overline{T}_{e2}} + \frac{\overline{\tilde{T}_{e2}(t)\tilde{N}_1(t)}}{\overline{T}_{e1}\overline{T}_{e2}}. \tag{7}$$

The particular case $\tilde{S}_1(t) = \tilde{S}_2(t) = \tilde{S}(t)$ defines the autocorrelation function. Whenever both signals are coming from the same plasma volume, one can assume $\tilde{T}_{e1} \approx \tilde{T}_{e2} = \tilde{T}_e$. Because the thermal noise is not correlated, the last three terms in the Eq. (7) can be neglected [5]:

$$R_{12}(0) = \frac{\overline{\tilde{T}_e^2(t)}}{\overline{T}_e^2}. \tag{7a}$$

From Eq. (7a), an expression for the root mean square (*rms*) value of the normalized temperature fluctuations can be obtained:

$$\frac{\sqrt{\overline{\tilde{T}_e^2(t)}}}{\overline{T}_e} = \sqrt{R_{12}(0)}. \tag{8}$$

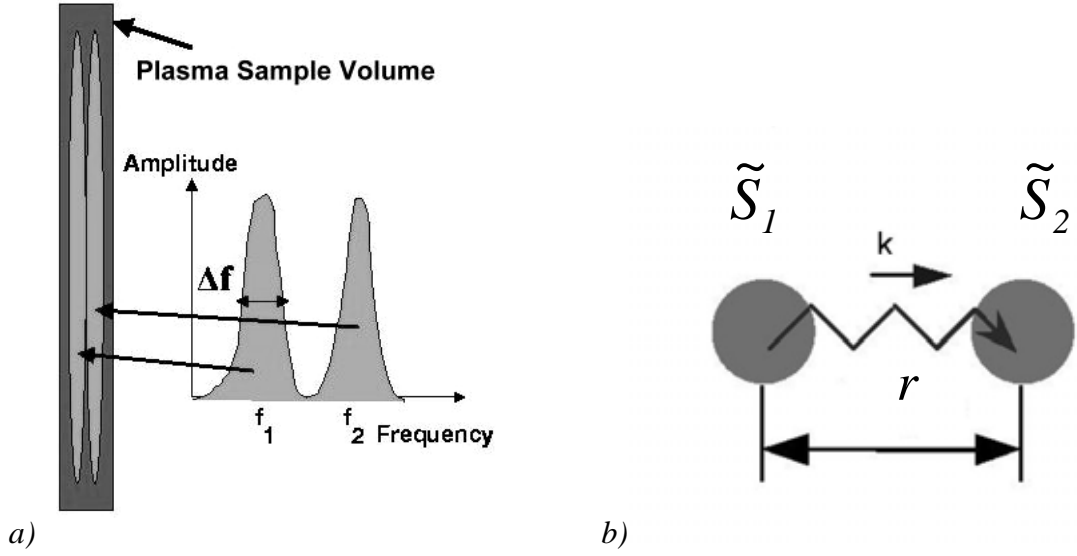

*a)* *b)*

***Figure 1***. *(a) - the thermal noise decorrelation with a single line of sight ECE system; (b) – cross-correlation between two spatially separated plasma volumes aids in determination of the dispersion relation k(f).*

The cross-correlation method does not require ECE signals to be absolutely calibrated. In many cases, besides the *rms* estimation of the fluctuations amplitude, it is useful to calculate the cross-spectral density spectrum for two discrete signals, which is the inverse Fourier transform of the cross-correlation function. The one-sided cross-spectral density for two ECE signals can be defined as follows:

$$CSD_{12} = |CSD_{12}(2\pi f)| e^{-j\varphi(2\pi f)}, \qquad (9)$$

where $|CSD_{12}(2\pi f)|$ is the amplitude of the cross-spectral density, and $\varphi(2\pi f)$ is the cross-phase between two signals. These quantities can be used to estimate the wavenumber of fluctuations and to derive the dispersion relation, as it is described by Eq. (5).

If broadband modes with a bandwidth $B_{BB}$ exist in the plasma, they cause a peak in the cross-correlation function that decays according the following time scale [5]:

$$\tau_{decay} = \frac{1}{B_{BB}} \sqrt{\frac{ln2}{\pi}}. \qquad (10)$$

## 3. Diagnostic set-up

The Tore Supra heterodyne radiometer [12] has been recently upgraded to have 1*GHz* spaced, 500 *MHz* bandwidth 32 measuring channels (Fig. 2). It is being used on the Tore Supra tokamak to measure the electron cyclotron emission in the frequency range 78-110 *GHz* for the first harmonic ordinary (O) mode ($\vec{E} \parallel \vec{B}, \vec{k} \perp \vec{B}$) and 94-126.5 *GHz* for the second harmonic extraordinary (X) mode ($\vec{E} \perp \vec{B}, \vec{k} \perp \vec{B}$). The radial resolution is essentially limited by ECE relativistic effects related to electron temperature and density

and not by the channels frequency spacing. The radiometer can act simultaneously in two modes:

1. *slow acquisition mode during all the plasma duration*: 32 channels 1 *ms* sampling without aliasing (bandwidth $B_{V1} = 400$ *Hz*);

2. *fast acquisition mode during time plasma windows triggered by plasma phenomenon*: 32 channels 10 *μs* sampling without aliasing (bandwidth $B_{V2} = 40$ *kHz*).

A precise absolute spectral calibration is performed outside the tokamak vacuum vessel by using a 600°C black body source. Using analytical formulas, post-pulse data processing takes routinely into account the total magnetic field and the Maxwellian relativistic radial shift to improve radial location estimate. These formulas are compatible with real time processing in order to use ECE data in feedback control loops.

To perform measurements of the electron temperature fluctuations, one radiometer channel is split into two (Fig. 3). On each of these two channels, an (IF) YIG filter with bandwidth around 100 *MHz* is introduced. Its central frequency is remotely monitored by a driver (designed at the IPP Forschungszentrum Jülich, Germany) between 6 and 18 *GHz*, allowing to shift the observation volume in the plasma radially. The IF filters, a Schottky diode detector and a video amplifier with 200 *kHz* bandwidth are placed inside the isolation box. The acquisition is done without aliasing effects.

It is important to mention that the ripple in Tore Supra reaches 7% at the edge leading to a mismatch between the iso-***B*** lines, which determine the localization of the investigated volume, and the field lines to which turbulence is perpendicular (Fig. 4). Consequently, for small distance between the two channels the same field line would cross the two volumes.

The minimum distance that can be reached without including this spurious effect is determined as:

$$\Delta R > w_{ECE} \tan(\alpha). \qquad (11)$$

The beam waist $w_{ECE}$ is equal to 5.4 cm and is taken to be the same for the two channels. The angle $\alpha$ is the difference between the curvature of the iso-***B*** and the field lines. The angle of the field lines is negligible with respect to that of the iso-***B*** lines and is thus neglected leading to en expression of $\alpha$ of the form:

$$\tan(\alpha) = N_c \sin(N_c \phi) \Delta B_\varphi. \qquad (12)$$

Here, $N_c$ is the number of toroidal field coils (equal to 18), $\phi$ is the ECE radiometer viewing angle with respect to the port axis ($\phi = 3.5$ degrees) and $\Delta B_\varphi$ is the magnetic ripple that varies between 0.18 (in the centre) and 0.93% (at $r/a = 0.4$), depending on the radial position. This leads to $\alpha$ between 1.6 and 8.5 degrees, respectively. Consequently, the contribution of turbulent fluctuations on the same flux surface to the two channels can be neglected for distances greater than 1.5 *mm* for measurements in the plasma centre and 8 *mm* at $r/a = 0.4$.

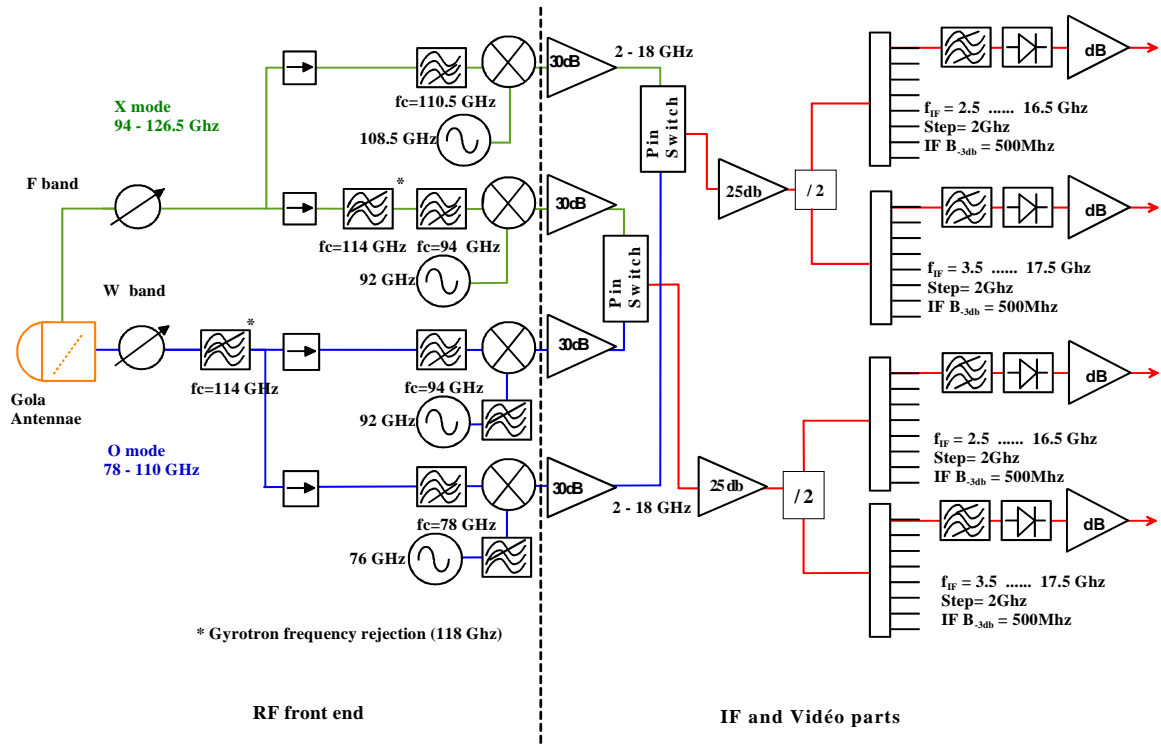

*Figure 2.* A principle scheme of the 32-channel heterodyne radiometer on Tore Supra. This figure is taken from [12] with kind permission of the authors.

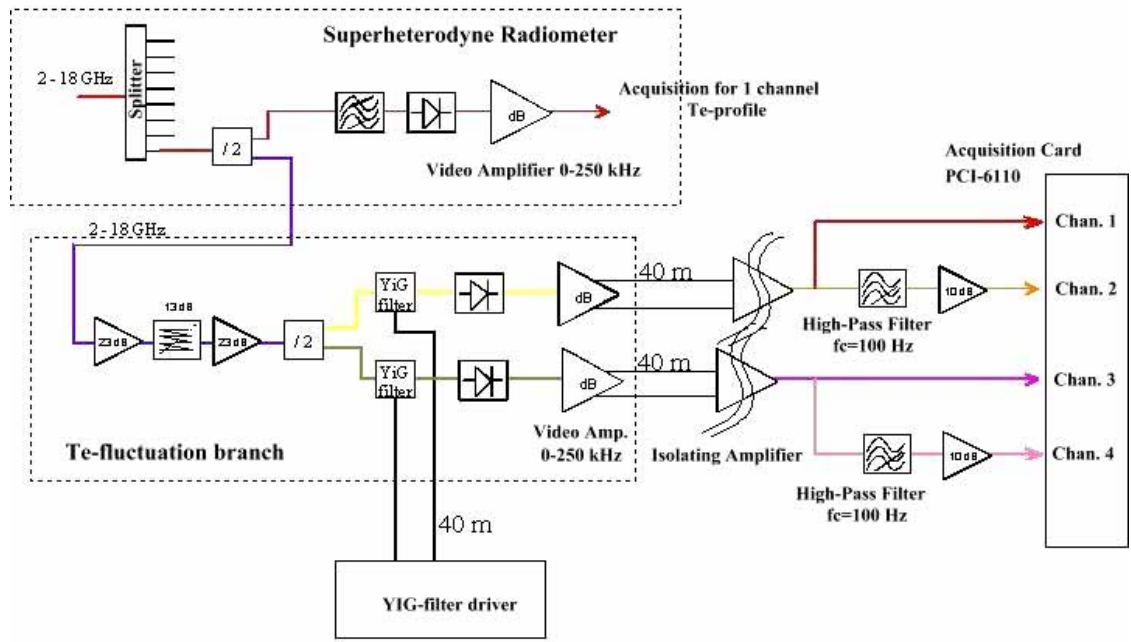

*Figure 3.* The principal electronic scheme to measure electron temperature fluctuations on Tore Supra.

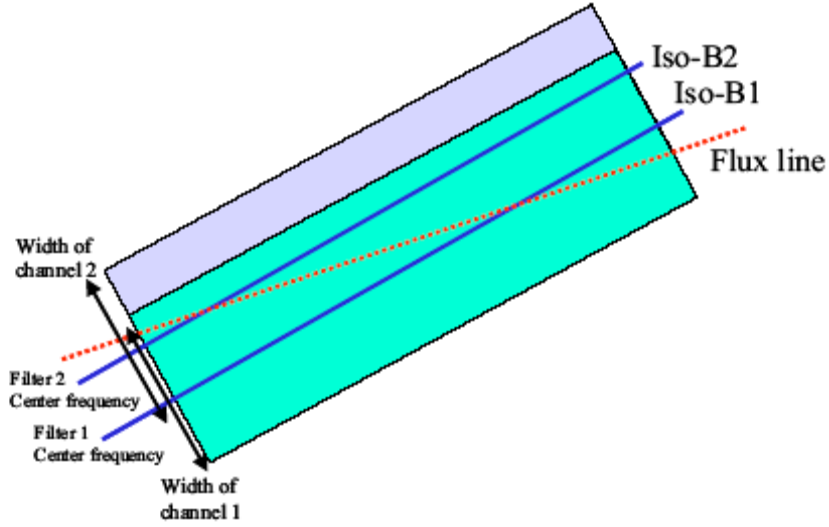

*Figure 4. A schematic view of the integration volume in the equatorial plane as well as two iso-**B** lines and one field line. As the distance between the two decreases, the mixing of the information comes from not only the width of the beams but also from the difference between the iso-**B** and the field lines.*

## 4. Estimation of the measurement error

*4.1 Signal to noise ratio for the Tore Supra radiometer*

ECE radiation coming from the plasma is attenuated by the waveguides losses and millimetric attenuator for total $A = 4\ dB + 17\ dB = 21\ dB$. The source temperature $T$ seen by the radiometer can be defined as follows:

$$T=\frac{T_e}{A'}+T_N, \qquad (13)$$

where $A'$ is about 100 (for 21 *dB* attenuation), and $T_N$ is the equivalent radiometer temperature noise.

The noise equivalent power (NEP) for one polarisation direction due to fluctuations of the intensity of the thermal radiation equals to [13]:

$$NEP=kT(B_{IF}B_v)^{\frac{1}{2}}, \qquad (14)$$

where $B_{IF}$, $B_v$ are IF (SSB) and video bandwidths, respectively. If one assumes $T_N$ to be about 10 000 K (due to the mixer and the first IF amplifier; the Shottky detector noise is negligible in case of strong IF amplification), the minimum detectable temperature difference $\Delta T_{min}$ is obtained when the black body radiated power is equal to the NEP (i.e. when signal to noise ratio is unity):

$$k\frac{\Delta T_{min}}{A'}B_{IF} = k(\frac{T_e}{A'}+T_N)(B_{IF}B_v)^{\frac{1}{2}}.\qquad(15)$$

If $T_e/A' \gg T_N$, Eq. (15) can be written as follows:

$$\frac{\Delta T_{min}}{T_e} = \sqrt{\frac{B_v}{B_{IF}}}.\qquad(15a)$$

For $B_v = 200\ kHz$ and $B_{IF} = 100\ MHz$, one gets the minimum detectable temperature ratio of about 4.4 %. If $T_e = 3\ keV$, $\Delta T_{min}$ is estimated to be 132 $eV$.

*4.2 Statistical noise level*

In order to reduce the statistical error below the coherent temperature fluctuation amplitude, a long sampling time is required. According to [11], the difference of $R_{12}$ (see Eq. (6)) (in case $\tilde{S}_1(t)$ and $\tilde{S}_2(t)$ consist of white (Gaussian) noise with a large bandwidth $B$) can be written as:

$$Var[R_{12}(\tau)] = \frac{1}{2BT}\left[R_{12}(0)R_{12}(0)+R_{12}^2(\tau)\right],\qquad(16)$$

where $T$ is the total integration time. For the present correlation ECE diagnostic on Tore Supra, the following ratio between the sampling frequency $M/T$ and the video band $B_V$ is valid: $2B_VT \approx M/3$, where $M$ is the total number of samples. For the cross-correlation function, one can write:

$$\Delta\left[\frac{\overline{\tilde{T}_e^2(t)}}{\overline{T}_e^2}\right] = \sqrt{Var[R_{12}(0)]} = \frac{\sqrt{3}}{\sqrt{M}}\frac{\left[\left(\overline{\tilde{T}_e^2}+\overline{\tilde{N}^2}\right)^2+\left(\overline{\tilde{T}_e^2}\right)^2\right]^{1/2}}{\overline{T}_e^2};\qquad(17)$$

$$\Delta\left[\frac{\overline{\tilde{T}_e^2(t)}}{\overline{T}_e^2}\right] = \frac{\sqrt{3}}{\sqrt{M}}\left[\frac{\overline{\tilde{T}_e^2(t)}}{\overline{T}_e}\right]\left[\left(1+\frac{\overline{\tilde{N}^2}}{\overline{\tilde{T}_e^2}}\right)^2+1\right]^{1/2};\qquad(18)$$

or, for fluctuations with the weak amplitude:

$$\Delta\left[\frac{\overline{\tilde{T}_e^2(t)}}{\overline{T}_e^2}\right] = 2\frac{\sqrt{\overline{\tilde{T}_e^2(t)}}}{\overline{T}_e}\Delta\left[\frac{\sqrt{\overline{\tilde{T}_e^2(t)}}}{\overline{T}_e}\right].\qquad(19)$$

From Eqs. (18) and (19), one finally gets:

$$\Delta\left[\frac{\sqrt{\overline{\tilde{T}_e^{\,2}(t)}}}{\overline{T}_e}\right]=\frac{\sqrt{3}}{2\sqrt{M}}\frac{\sqrt{\overline{\tilde{T}_e^{\,2}(t)}}}{\overline{T}_e}\left[\left(1+\frac{\overline{\tilde{N}^{\,2}}}{\overline{\tilde{T}_e^{\,2}}}\right)^2+1\right]^{1/2}. \qquad (20)$$

It can be seen that the statistical error depends on $1/\sqrt{M}$ and not on $1/\sqrt[4]{M}$, as can be found elsewhere in the literature [2, 5]. For example, to resolve the fluctuation amplitude of 0.2%, it is necessary to have $M = 10^6$ samples to get the error level of 0.1%. For the fluctuation amplitude of 0.1%, $3\times 10^6$ samples are required for the same error level of 0.1%. Error estimation (standard deviation) for real signal sequences is shown in Fig. 5 and deviates slightly from both scaling laws.

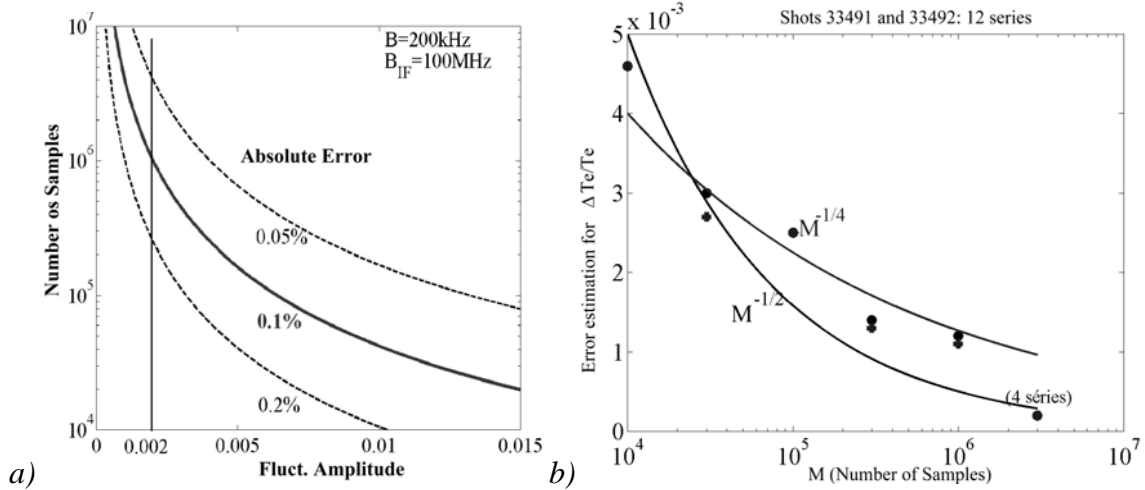

*Figure 5. The number of samples versus relative temperature fluctuations amplitude to resolve the desired absolute error (a), and error estimation (standard deviation) for 12 series of real correlation ECE data with different number of samples (b). For (b), ECE signals from two successive Tore Supra shots have been taken. Solid lines represent $1/\sqrt{M}$ and $1/\sqrt[4]{M}$ scaling laws for the error estimation.*

One additional, though very important, note has to be mentioned. The sampling rate of the diagnostic is 1 *MHz*, however, the video bandwidth is limited by 200 *kHz*. Therefore, an oversampling effect is present in correlation ECE measurements on Tore Supra. Because of this, the high-frequency tail of the spectra (> 200 *kHz*) cannot be used to analyse the spectral characteristics of fluctuations. Precautions should also be taken when looking to the cross-correlation functions at time lags smaller than $5\times 10^{-6}$ *s*. Resampling the input ECE signals at 250 *kHz* (or smaller frequency) is desirable.

## 5. First measurements

Depending on current in the poloidal coils and, hence, on the toroidal magnetic field in the plasma centre, a few measuring scenarios are possible for the correlation ECE on Tore Supra. Figure 6 gives the radial region of observation by the correlation ECE as a function of the current in the poloidal coils: $I_{coils} = 1100\ A$ corresponds to the central toroidal field of about 3.4 *T*, and $I_{coils} = 1300\ A$ corresponds to $B_T = 4\ T$. The value of the Shafranov shift depends on heating regime and typically varies between 5 and 12 *cm*.

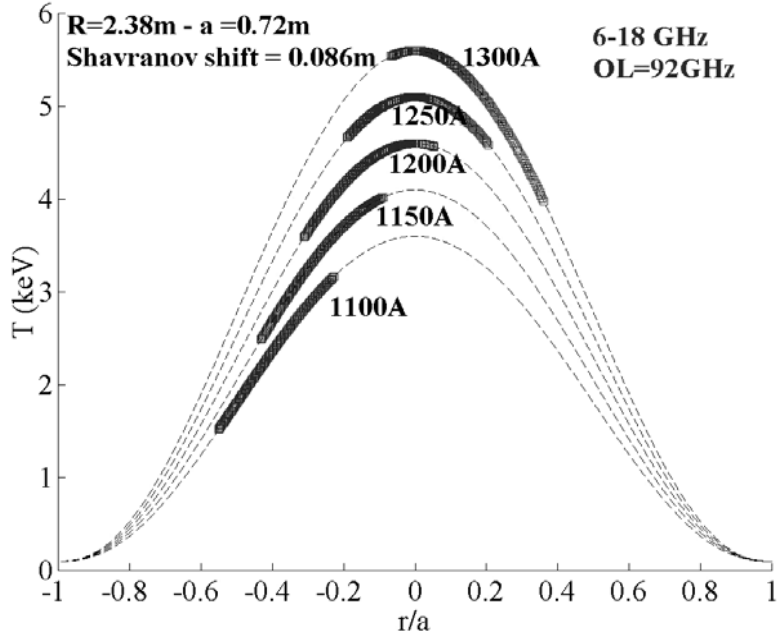

*Figure 6.* A radial range of correlation ECE measurements for different values of the poloidal coil current on Tore Supra.

*5.1 Low-frequency MHD test studies by 32-channel (non-correlation) ECE radiometer*

In order to test the theoretical principles described in Section 2, determination of the well known harmonic component, such as the $m/n = 1/1$ precursor to sawteeth, has been done by means of cross-correlation between two channels of the 32-channel ECE radiometer (non-correlation set-up). Figure 7 shows a simple example of the power spectral density estimation by FFT applied directly to a typical sawtooth signal with the $m/n = 1/1$ magnetohydrodynamic (MHD) precursor activity. Figure 8 shows the cross-correlation between two neighbouring channels of the present 32-channels heterodyne ECE radiometer (sampling rate 83 kHz) for the same shot but for filtered signals, in order to get rid of the sawtooth rise.

It can be seen that the shape of the signal can influence correlation functions and resulting spectra. If the amplitude of fluctuations (and, therefore, correlation between channels) is strong, like in the example shown in Figs. 7 and 8, the real structure in the plasma becomes visible even without smoothing the input signals. However, to uncover a presence of some broadband high-frequency modes that may exist in the plasma but hardly detectable because of the noise, filtering or smoothing of the input signals is desirable, so the low-frequency step-like components (such as sawteeth) and high-frequency noise contribution would not influence the tail of the spectra.

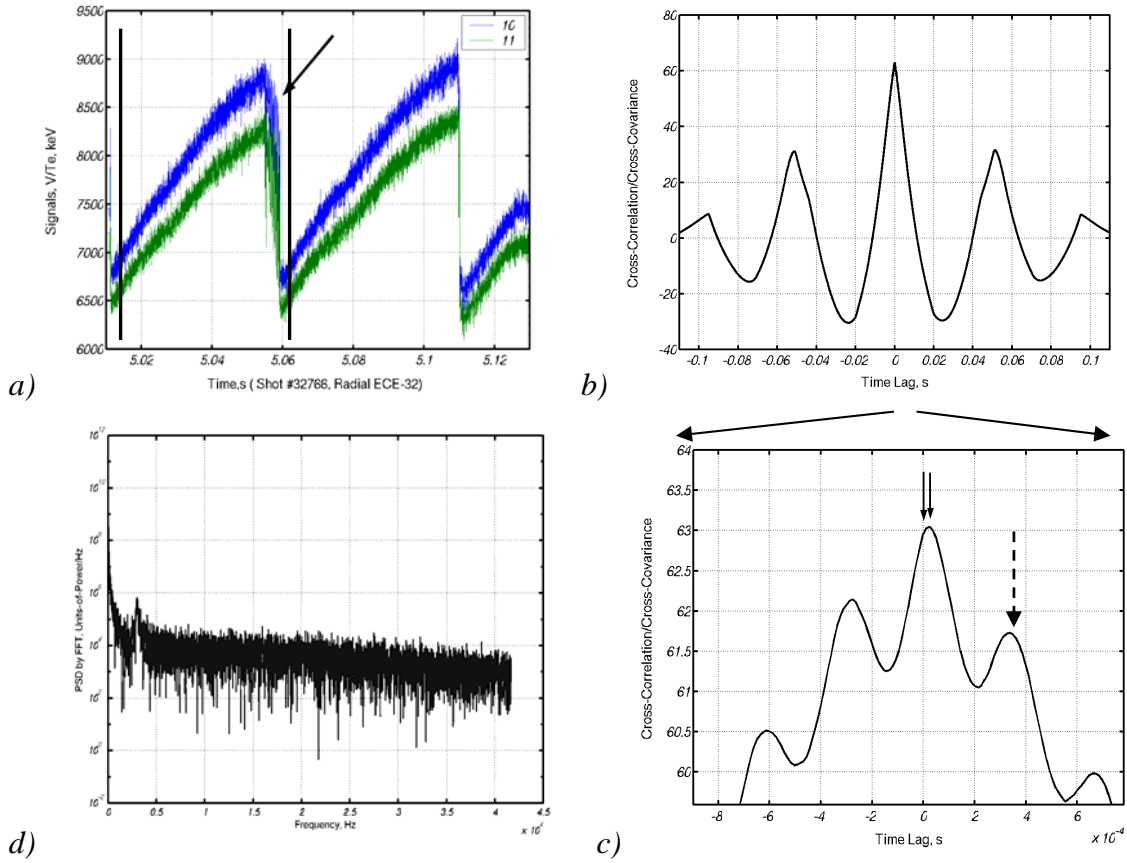

*Figure 7. A power spectral density estimation (d) for a single ECE signal during sawtooth activity with the strong m = 1 precursor (a; selected signal is indicated by the black arrow). Vertical lines in (a) show the time interval for which the PSD has been estimated. Typical cross-correlation functions for the whole time window in (a) are given in (b) and (c), and depict a strong influence of sawteeth. Black solid arrows in (c) indicate the propagation of the m = 1 precursor between two plasma sample volumes. Dashed arrow gives the time lag between two successive precursor oscillation of about $3.3 \times 10^{-4}$ s, from which the rotation frequency of about 3 kHz is deduced.*

5.2 Broadband mode identification by correlation ECE

It is known [3, 14] that the gradient of the electron temperature is one of the energy source of the turbulent fluctuations. A radial scan at the HFS on Tore Supra has been performed between $r/a$ = (-0.6) – (-0.1) for the central magnetic field $B_T$ = 3.3 T. A radial separation between two correlation ECE channels of about 1 *cm* has been chosen. An integration time of 2 *s* has been used. A qualitative comparison between cross-power spectral densities for two extreme cases, one inside the sawtooth inversion radius at $r/a$ = (- 0.18), and one in the $T_e$-gradient region at $r/a$ = (- 0.43), has shown a noticeable difference in the frequency range of 20 – 150 *kHz* (see Fig. 9). The origin of the "bump" for the CSD in the gradient region is not clear yet and may be either an effect of the Doppler shift due to the poloidal plasma rotation transposing spectral components to the higher frequencies, or a footprint of the broadband turbulent mode of electro-magnetic nature. The latter is more likely, because the "bump" feature has much weaker amplitude at $|r/a| > 0.5$, whilst one would expect the Doppler effect to persist with the higher amplitude with increased value of $|r/a|$. Similar observations have been done at the LFS, too.

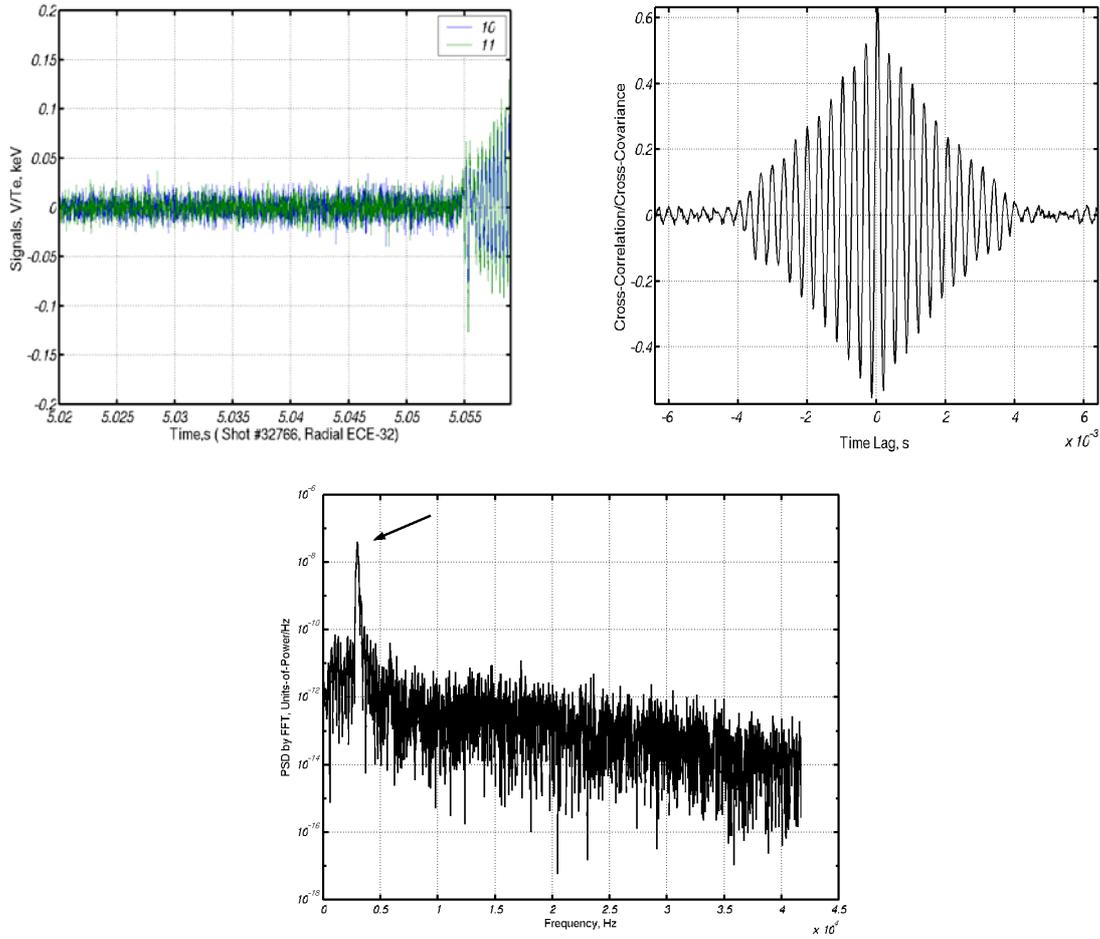

*Figure 8. Example of cross-correlation function and cross-power spectral density for sawtooth-removed ECE signals. Black arrows on the CSDs give the precise rotation frequency of about 3 kHz for the precursor. Note that the CSD slope at frequencies below 2kHz is absent in case of filtered ECE signals.*

From the cross-correlation functions in Fig. 9, it is seen that the broadband structure with the maximum at $1.9 \times 10^{-5}$ s is superimposed on the low-frequency MHD components with the bandwidth below 10 $kHz$. From Fig. 10, an estimation of the relative fluctuation amplitude for a statistical trustable time (and frequency) interval can be made: for $r/a = (-0.43)$, $\tilde{T}_e/T_e$ is about 0.35% for the broadband structure. For the shot with $r/a = (-0.18)$, the fluctuation level is much smaller and well below the noise level (most likely caused by the Schottky diode detector). Another example is shown in Fig. 10(c), in which correlation ECE monitors the plasma center, well inside the sawtooth inversion radius. It can be seen that, although the noise component is much weaker than for the shot shown in Fig. 9(d), the "real" fluctuations are significantly reduced in their amplitude, compared to the $T_e$-gradient region, and estimated to be about 0.2% .

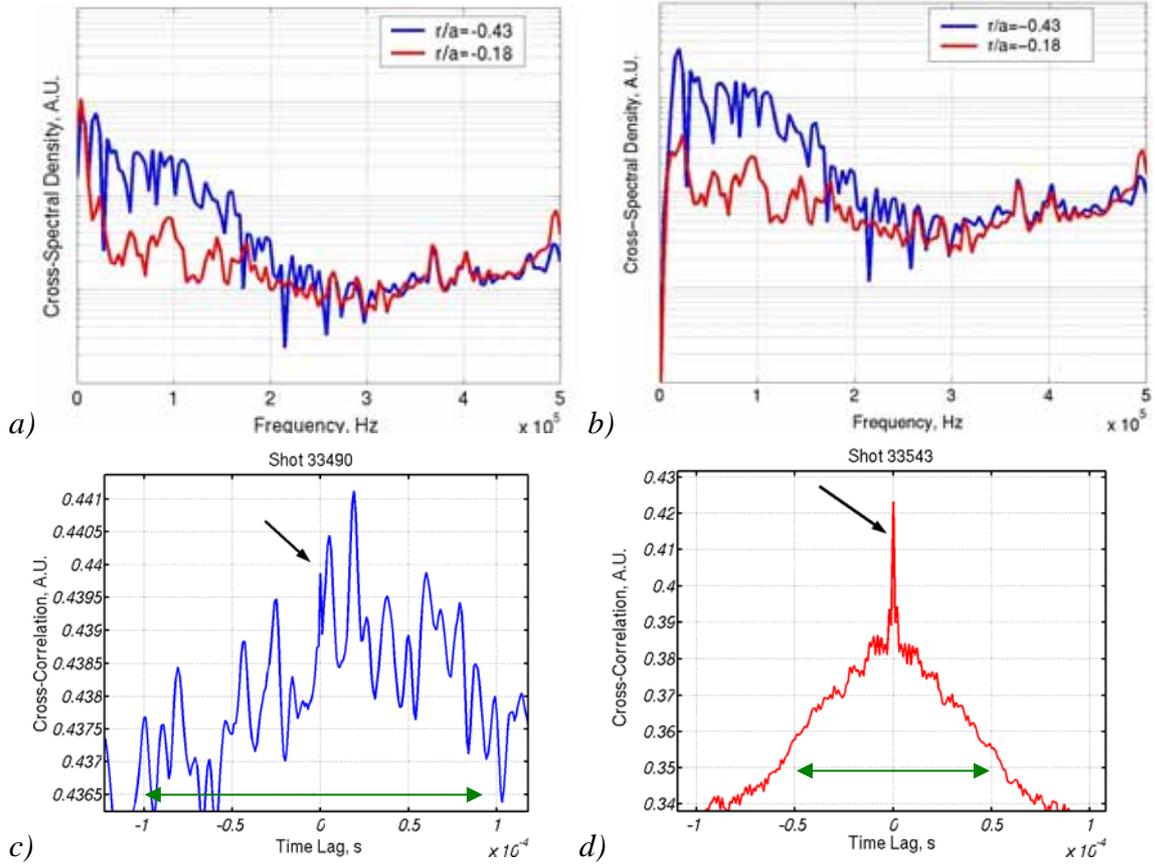

*Figure 9. Cross-spectral densities (in arbitrary units) for correlation ECE at different radial location: blue – in the $T_e$-gradient region at the HFS, red – inside the sawtooth inversion radius at the HFS. Plot (a) shows the CSD for ECE signals have been detrended before FFT calculations, in order to prevent low-frequency components due to MHD and high-frequency components due to the instrumental noise to influence the level (offset) of the spectra; plot (b) is for the CSD out of high-pass filtered (< 20 kHz) ECE signals (to eliminate the low-frequency MHD effects totally). Cross-correlation function for the case of r/a = -0.43 (c) shows a presence of high-frequency components that are enhanced in the amplitude, compared to the noise peak (black arrow) at zero time lag; for the case of r/a = -0.18 (d), the noise component is dominant. Green horizontal lines in (c) and (d) shows an offprint of the low-frequency mode of MHD origin with a bandwidth below 10 kHz, on which high-frequency components are superimposed.*

From the phase shift of the broadband structure, the radial propagation velocity is found to be about 500 *m/s*. More experiments dedicated to study broadband high-frequency modes are planned in upcoming campaigns on Tore Supra.

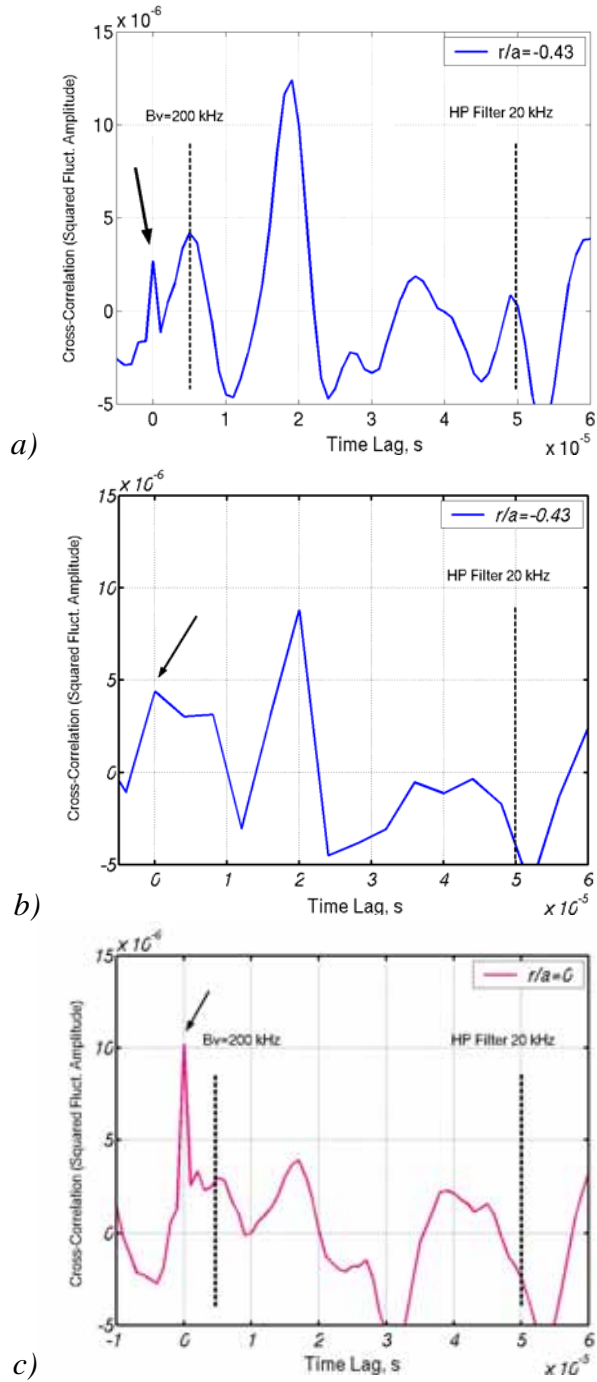

*Figure 10.* A (biased) cross-correlation function for high-pass filtered (20 kHz) and mean-normalized ECE signals (1 MHz sampling rate) at r/a = -0.43 (a), and for the same signals but resampled at 250 kHz (b). Vertical dotted lines show a statistically trustable time interval: $5 \times 10^{-5}$ s (and above) correspond to 20 kHz (and lower) in frequency field both in (a) and (b), and $5 \times 10^{-6}$ s correspond to 200 kHz (limitation by the video detector bandwidth in (a)). Noise peak at zero time lag (and, therefore, above 200 kHz) is shown by black arrows. Plot (c) is for the case in which correlation ECE monitors plasma inside the inversion radius, near the plasma center. For this shot (#33505), the noise component (black arrow) is much smaller, compared to the shot presented in Fig. 9(d).

## 6. Summary and future plans

The feasibility of electron temperature fluctuation measurements on Tore Supra has been investigated by means of radial correlation ECE diagnostic. Though the range of these measurements was limited by the toroidal magnetic field (and also by the restricted possibility to change the central frequency of the YIG filter), and only a few experiments were conducted so far, first observations of the broadband frequency structure have been done in the $T_e$-gradient region, compared to the plasma core inside the sawtooth inversion radius. From the collected data, there is no possibility yet to derive the dispersion relation and to identify the origin of this structure. In upcoming experimental campaigns, the following studies by means of correlation ECE radiometer are proposed on Tore Supra:
1. to perform a detailed radial scan both on the LFS and the HFS to identify the radial extent, correlation length and decorrelation time of the broadband mode and, thus, to derive its wavenumbers;
2. to study the response of turbulence amplitude at different values of the plasma density (e.g. at different collisionality);
3. to verify the results reported in [7] that turbulence is separated inside and outside the transport barrier, in particular with respect to the rational $q$ surfaces;
4. to compare the temperature fluctuation measurements with the density fluctuation measurements to obtain a rather comprehensive picture of turbulence in the plasma core;

In the year 2005, a further upgrade of the correlation ECE radiometer is planned. The future diagnostic will have several tuneable channel sets allowing detailed studies of temperature fluctuations over a wider radial range.